\documentclass[prd,aps,twocolumn,nofootinbib,showpacs,amsmath,superscriptaddress]{revtex4-2}

\usepackage{graphicx}
\usepackage{epstopdf}
\usepackage{bm}
\usepackage{epsfig}
\usepackage{graphics}
\usepackage{xspace}
\usepackage{amssymb}
\usepackage{amsmath}
\usepackage{xcolor}
\usepackage[utf8]{inputenc}
\usepackage[normalem]{ulem}
\usepackage{stmaryrd}

\def\OMIT#1{}

\newcommand{\bea}{\begin{eqnarray}}
\newcommand{\eea}{\end{eqnarray}}

\newcommand{\gsim}{\mathrel{\rlap{\lower4pt\hbox{\hskip1pt$\sim$}}\raise1pt\hbox{$>$}}}

\newcommand{\be}{\begin{equation}}
\newcommand{\ee}{\end{equation}}

\begin{document}
\title{Probing charged lepton flavor violation with a positron beam at CEBAF(JLAB)}
\author{Yulia Furletova}

\affiliation{Thomas Jefferson National Accelerator Facility}

\author{Sonny Mantry}
\affiliation{Department of Physics and Astronomy, 
                   University of North Georgia,
                   Dahlonega, GA 30597, USA}

\begin{abstract}

The addition of a high intensity 11 GeV polarized positron beam at the Continuous Electron Beam Accelerator Facility (CEBAF) at JLAB would allow for a search of Charged Lepton Flavor Violation (CLFV) via the process $e^+N \rightarrow \mu^+ X$. The proposed Solenoidal Large Intensity Detector (SoLID) spectrometer, in the configuration with muon chambers, would be ideal for such CLFV searches. Various new physics scenarios, including the phenomenologically convenient Leptoquark (LQ) framework, predict CLFV rates that are within reach of current or planned experiments.  A positron beam with instantaneous luminosity, ${\cal L}\sim 10^{38}$ cm$^{-2}$s$^{-1}$, could improve on existing HERA limits by two or three orders of magnitude. The availability of positron beam polarization would also allow for distentangling CLFV effects mediated by left-handed vs. right-handed LQs. 
\end{abstract}

\maketitle

\section{Introduction}
\label{intro}

\begin{figure}[hbt]
\includegraphics[width=0.5\textwidth]{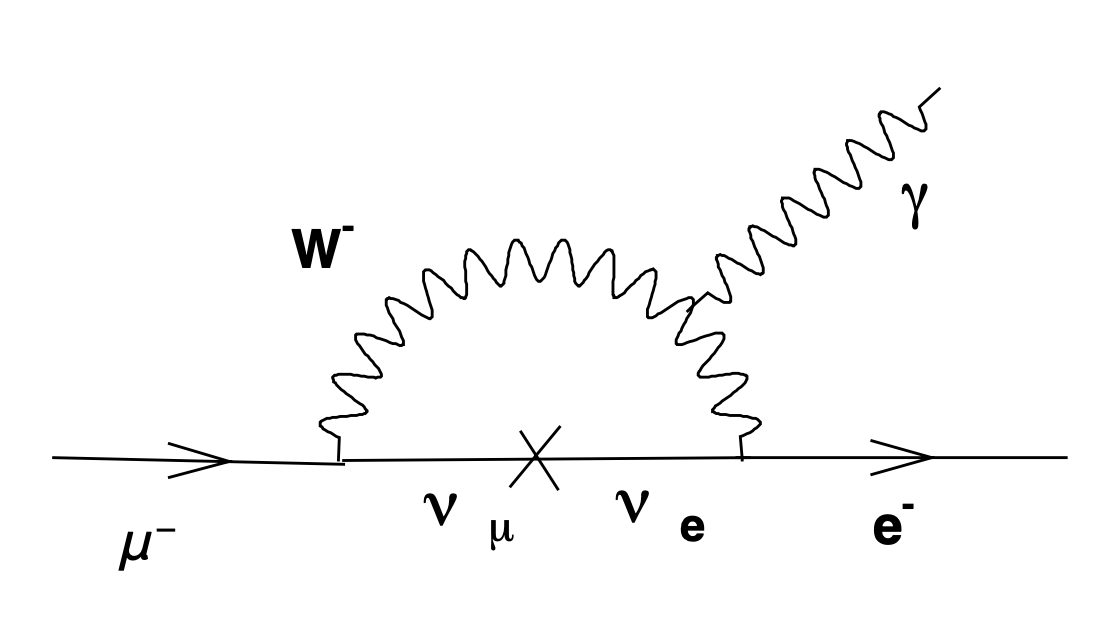}
\caption{ The one-loop CLFV process, $\mu\to e\gamma$, mediated via lepton flavor violation in the neutrino sector.}
\label{fig:muegamma}
\end{figure}

The discovery of neutrino oscillations gave conclusive evidence that lepton flavor is not a conserved quantity. However, the observation of lepton flavor violation in the charged lepton sector has eluded all experimental seaches to date. In fact, the non-zero mass of neutrinos predicts the existence of charged lepton flavor violating (CLFV) processes, such as $\mu \rightarrow e \gamma$, through loop induced mechanisms, as seen in Fig.~\ref{fig:muegamma}. 
However, the smallness of the neutrino masses makes this process highly suppressed with a branching fraction of Br($\mu \rightarrow e \gamma$ ) $< 10^{-54} $\cite{Baldini}, far beyond the reach of any current or planned experiments. 

\begin{figure}[hbt]
\includegraphics[scale=0.9]{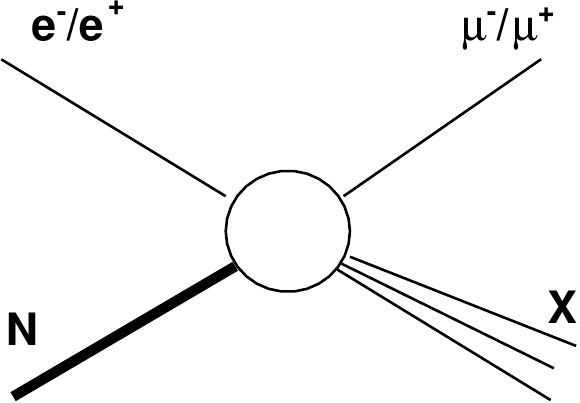}
\caption{Schematic of the CLFV DIS process  $e^\pm N \rightarrow \mu^\pm X$.}
\label{fig:EN}
\end{figure}

However, many beyond the Standard Model (BSM) scenarios~\cite{PDG:2020} predict significantly higher CLFV rates that are within reach of current or future planned experiments. A variety of experiments across the energy spectrum have searched for and set limits on CLFV processes that involve transitions between the electron and the muon. These include searches for muon decays $\mu^-\to e^-\gamma$ (MEG experiment  ~\cite{Mu2eBi})  and $\mu^-\to e^-e^-e^+$ (Mu3e experiment  ~\cite{MU3e}), the $\mu-e$ conversion process $\mu^- + A(Z,N) \to e^- + A(Z,N)$ (SINDRUM ~\cite {SINDRUM-2} and COMET ~\cite{COMET} experiments), and the Deep Inelastic Scattering (DIS) process $e^\pm N \to \mu^\pm X$ ~\cite{HERA-ZEUS}.  The most stringent limits come from MEG~\cite{MEG-II}, Br$(\mu \to e\gamma) < 4.2 \times 10^{-13}$, and SINDRUM II~\cite{SINDRUM}, CR($\mu-e, Au)< 7.0\times 10^{-13}$. The H1~\cite{H1:limit} and ZEUS~\cite{HERA-ZEUS} collaborations at HERA have also set limits through searches for the CLFV DIS process $e^\pm N \to \mu^\pm X$, seen in Fig.~\ref{fig:EN}. While some of these CLFV limits are stronger than others, each can provide complementary information since they can probe different CLFV mechanisms in different types of processes. Furthermore, CLFV searches involving muons could have new significance in light of the recently observed muon anomalies such as the muon g-2 measurement~\cite{Bennett_2006,Abi_2021} and the B-decay ratios $R_{K^{(*)}} = \text{Br}(B\to K^{(*)}\mu^+\mu^-)/\text{Br}(B\to K^{(*)}e^+e^-)$~\cite{Aaij_2017,Aaij:2021vac} and $R_{D^{(*)}} = \text{Br}(B\to D^{(*)}\tau \bar{\nu}_\tau)/\text{Br}(B\to D^{(*)}e(\mu) \bar{\nu}_{e(\mu)})$~\cite{Aaij_2018,Abdesselam:2019dgh}.

Here we explore the possibility of studying CLFV with a polarized positron beam~\cite{Accardi:2020swt}  at the Continuous Electron Beam Accelerator Facility (CEBAF) at JLAB in the DIS  process: 

\begin{eqnarray}
e^+ + N\to \mu^+ + X. 
\end{eqnarray}

\section{Charged Lepton Flavor Violation at CEBAF}
\label{sec:1}



A high intensity positron beam~\cite{Accardi:2020swt} at the CEBAF at JLAB can search for the CLFV process $e^+N \rightarrow \mu^+ X$. The 11 GeV polarized positron beam will impinge on a proton target at rest, corresponding to a center of mass energy, $\sqrt{s} \sim 4.5$ GeV.  In spite of the relatively small center of mass energy, the high luminosity, ${\cal L}\sim 10^{36-39}$ cm$^{-2}$s$^{-1}$, will allow for significant improvement on existing limits from HERA~\cite{H1:limit,HERA:ZEUS}.  


The experiment should be equipped with detectors, which could provide a trigger for muons (for example, muon chambers or a tagger after the hadron-absorber), as well as a good tracker and, if possible, a vertex detector, to minimize background from pion-decays. CLFV events have a similar topology to DIS events where the scattered electron is replaced by muon. The selection should be  based on events which do not have electrons in the final state, but instead have a clear evidence of a muon track pointing to the vertex.

The proposed SoLID spectrometer(Solenoidal Large Intensity Detector) ~\cite{chen2014white} will meet the above requirements. This high-luminosity and high-acceptance detector has been proposed for the JLAB  12 GeV program, and will be able to handle the expected  high
luminosity, ${\cal L} \sim 10^{36}$ - $10^{39}$ cm$^{-2}s^{-1}$. In addition, SoLID can carry out measurements not only using high intensity unpolarized or polarized lepton beams, but also  unpolarized or polarized nuclear targets, which will be  important for distinguishing between different CLFV mechanisms~\cite{Taxil_2000}. 

\begin{figure}[hbt]
\includegraphics[width=0.5\textwidth]{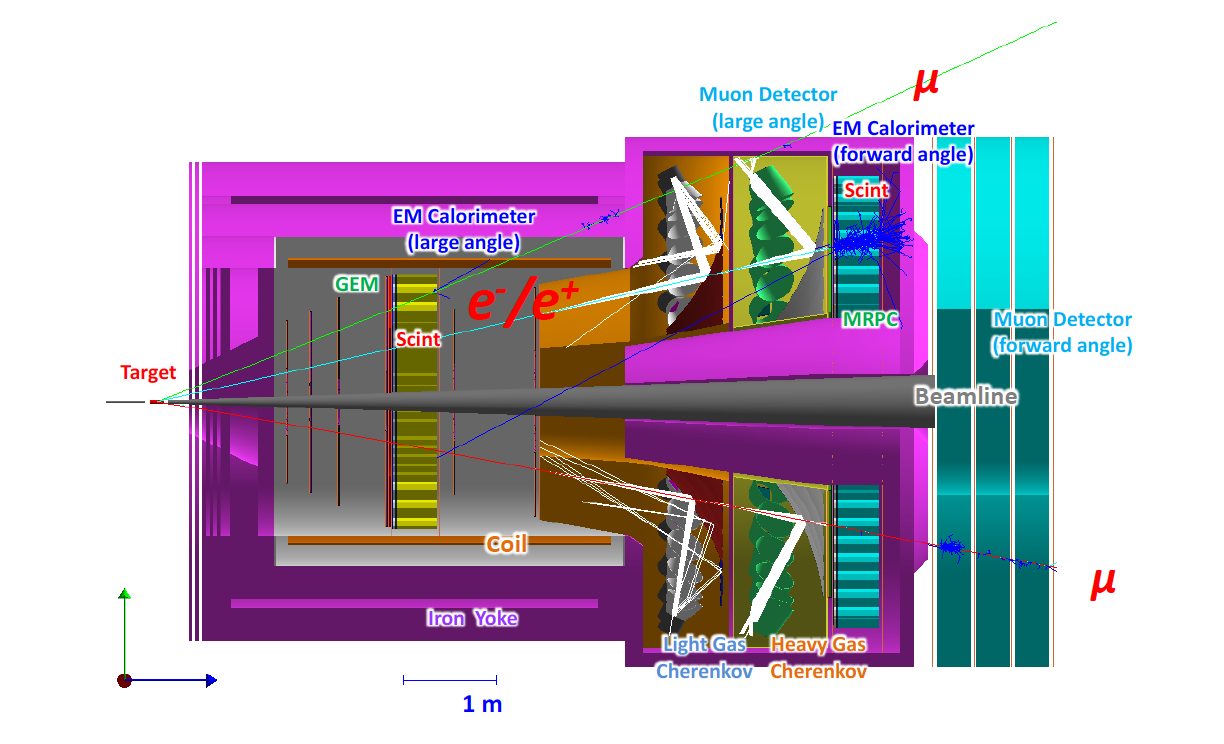}
\caption{The SoLID $J/\Psi$ configuration with muon detectors~\cite{zhao2021double}. Other sub-detectors are labeled.}
\label{fig:SOLID}
\end{figure}

The SoLID experiment  will run in  different detector  configurations~\cite{chen2014white}, such as the $J/\psi$ production, Parity-Violating Deep Inelastic Scattering (PVDIS),  or the dedicated Double Deeply Virtual Compton Scattering (DDVCS) configuration. For CLFV measurements $J/\psi$ and DDVCS setups will be preferable, since both or them will be equipped with muon chambers. Fig.~\ref{fig:SOLID} shows the $J/\psi$ setup with muon chambers. The CLFV experimental program could run simuntatiously with the other approved experiments, since it will not require any additional hardware equipment. In the $J/\psi$ configuration,  the SoLID spectrometer will be equipped with large-angle and  a forward-angle  muon detectors. In addition, high resolution Gas Electron Multiplier (GEM) chambers, Cherenkov detectors, and Calorimeters will help muon momentum reconstruction and identification. The expected  muon detection efficiency in this setup is about $70\%$ for a single muon ~\cite{zhao2021double}.  

The SoLID experiment will have an acceptance in the polar angle, $\theta$, in the range of 8$^o$ to 24$^o$ and 22$^o$ to 35$^o$ for the SIDIS and PVDIS configurations, respectively, and full-$2\pi$ acceptance in the azimuthal angle $\phi$. This is typical for fixed target configurations where most of the cross section lies in the forward region due to the overall kinematic boost of the 11 GeV electron incident of the stationary proton.

Muon backgrounds must be suppressed or under control in order to extract bounds on the $e^+\to \mu^+$ CLFV process. Due to the compact size of the detector, the typical decay length of pions is much bigger than the distance to the detector from their production vertex. The survival probability of a pion at a distance $L$ away from its production vertex is given by~\cite{jenkins}
\begin{eqnarray}
\nonumber \\
P(L) = e^{-L/\lambda_D^\pi}, \qquad \lambda_D^\pi = \frac{p_\pi}{m_\pi c} c\tau, \\ \nonumber
\end{eqnarray}
where $\lambda_D^\pi$ is pion decay length and $\tau=26$ ns is the mean-life of the pion in its rest frame. For example, at SoLID, the pions will be produced with typical momenta, $p_\pi$, in the range of 1 GeV to 7 GeV~\cite{solidCDR}. This corresponds to a range in the decay length of about 56m to 390m. This range of decay lengths are to be compared with the distance of ~5m corresponding to the overall detector dimensions combined with its promiximity to the pion production vertex. This results in a pion survival probablity range between 91$\%$ and 99$\%$ at a distance of 5m from the pion production vertex. Thus, the muon background from pion decays is highly suppressed at SoLID compared to other fixed target experiments with large or non-compact detectors.

In order to further suppress the muon background from pion decays and or cosmic rays, it is important to have high precision charged particle tracking. Such tracking information will be used to recontruct the charged particle trajectories and their production vertices. This allow for separating any signal muons produced at the CLFV verter from the background muons coming from pion decays. In addition, the low center of mass energy $\sqrt{s}\sim$ 4.5 GeV implies there will no muon backgrounds from the decays of open charm or bottom mesons. However, there can be muon backgrounds from the production of J/Psi, via the strong interaction pair production of $c\bar{c}$, which can be easily rejected by tracking the resulting muon pair back to the J/Psi decay vertex. The SoLID experiment will have the capability for the needed charged particle tracking to reject muon backgrounds. In particular, it will have a tracking spatial resolution of 100 microns, allowing for a precise reconstruction of the muon decay vertices~\cite{solidCDR}.

\section{Leptoquark Mediated CLFV}
\label{sec:LQ}

It becomes convenient to study CLFV in the Leptoquark (LQ) scenario in which the CLFV DIS processes $e^\pm\to \mu^\pm +X$ can be mediated at tree-level.  LQs are color triplet bosons that mediate transitions between quarks and leptons and carry both baryon number and lepton number. As seen in Tables.~\ref{tab:F0} and \ref{tab:F2}, according to the Buchm\"uller, R\"uckl and Wyler classification~\cite{BRW:theory}, there are 14 different types of LQs characterized by their spin (scalar or vector), fermion number F=3B+L (0 or $\pm 2$), chiral couplings to leptons (left-handed or right-handed), $SU(2)_L$ representation (singlet, doublet, triplet), and $U(1)_Y$ hypercharge. 

The  $SU(3)_C \times SU(2)_L \times U(1)_Y$ invariant and renormalizable interactions are given by the Lagrangian for $F=0$ and $|F|=2$ LQs as follows:

\begin{eqnarray}
\label{eq:F0Lag}
{\cal L}_{F=0} &=& h_{1/2}^L \bar{u}_R \ell_L S^L_{1/2} + h^R_{1/2} \bar{q}_L \epsilon e_R S^R_{1/2}  +\tilde{h}^L_{1/2} \bar{d}_R \ell_{L} \tilde{S}^{L}_{1/2} \nonumber \\ 
&+& h_0^L\bar{q}_L\gamma_\mu \ell_L V_0^{L\mu} +h_0^R  \bar{d}_R \gamma_\mu e_R V_0^{R\mu} \nonumber \\
&+& \tilde{h}_0^R \bar{u}_R \gamma_\mu e_R \tilde{V}_0^{R\mu}
+ h_1^L \bar{q}_L \gamma_\mu \vec{\tau}\ell_L \cdot \vec{V}_1^{L\mu} + h.c. , \\ \nonumber
\end{eqnarray}

\begin{eqnarray}
\label{eq:F2Lag}
{\cal L}_{|F|=2} &=& g_0^L \bar{q}_L^c \epsilon \ell_LS^L_0 + g_0^R \bar{u}^c_R e_R S_0^R + \tilde{g}_0^R \bar{d}^c_R e_R \tilde{S}_0^R \nonumber \\
&+& g_1^L \bar{q}_L^c\epsilon \vec{\tau}\ell_L \cdot \vec{S}_1^{L} + g_{1/2}^L \bar{d}_R^c \gamma_\mu \ell_L V_{1/2}^{L\mu} \nonumber \\
&+&g_{1/2}^R \bar{q}_L^c \gamma_\mu e_R V_{1/2}^{R\mu} + \tilde{g}_{1/2} \bar{u}_R^c \gamma_\mu \ell_L \tilde{V}_{1/2}^{L\mu} + h.c. \\ \nonumber
\end{eqnarray}

\begin{figure}[hbt]
\centering
\includegraphics[width=0.5\textwidth]{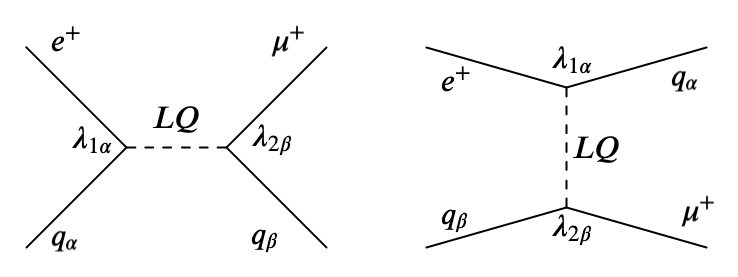}
\caption{The $e^+N\longrightarrow \mu^+X$ CLFV process mediated by the tree-level exchange of LQ states in the $s$ and $u$ channels.}
\label{fig:LQtreelevel}
\end{figure}
\begin{table}[]
{\scriptsize %
    \centering
    \begin{tabular}{|c|c|c|c|c|}
      \hline 
        Type  &  J & Q & s-channel process&coupling\\
 
        \hline 
        $S^L_0$ & 0  & -1/3 
        &  
        $e^-_L  u_L \rightarrow$ 
        \begin{tabular}{c} $l^- u $ \\ $\nu _l d$\end{tabular}
        
        & \begin{tabular}{c} $\lambda _L$ \\ -$\lambda _L$\end{tabular} \\
        \hline 
        $S^R_0$ & 0  & -1/3 
        &   $e^-_R u_R \rightarrow  \it{l}^- u$
        &  $\lambda _R$  \\
         \hline 
        $\tilde{S}^R_0$ & 0  & -4/3 
        &   $e^-_R d_R \rightarrow \it{l}^- d$
        &  $\lambda _R$  \\ 
        
       \hline 
        $S^L_1$ & 0 & \begin{tabular}{c} -1/3   \\ \\ -4/3\end{tabular} 
        &  
        \begin{tabular}{c} 
        $e^-_L u_L \rightarrow $ 
        \begin{tabular}{c} $l^- u $ \\ $\nu _l d$\end{tabular} \\
        $e^-_L d_L \rightarrow \it{l}^- d$
        \end{tabular} 
        
        & \begin{tabular}{c} -$\lambda _L$ \\ -$\lambda$ \\ -$\sqrt{2} \lambda_L$\end{tabular} \\

        \hline 
        \hline 
        $V^L_{1/2}$ & 1  & -4/3 
        & $e^-_L d_R \rightarrow  \it{l}^- d$
        & $\lambda _L$  \\ 
        
        \hline 
         $V^L_{1/2}$ & 1  &  \begin{tabular}{c}-1/3  \\ -4/3 \end{tabular}
        & \begin{tabular}{c} $e^-_R u_L \rightarrow  \it{l}^- u$ 
        \\  $e^-_R d_L \rightarrow  \it{l}^- d$  \end{tabular}
        & \begin{tabular}{c}  $\lambda _R  $
        \\   $\lambda _R$  \end{tabular} \\
        
      
      \hline 
        $\tilde{V}^L_{1/2}$ & 1  & -1/3 
        &   $e^-_L u_R \rightarrow  \it{l}^- u$
        &  $\lambda _L$   \\ 
       \hline
       
    \end{tabular}
    \caption{ The  $|$F$|$ = 2 leptoquarks in the Buchm\"uller-R\"uckl-Wyler classification.  For $|$F$|$ = 2  leptoquarks, the s-channel process dominates with an electron beam due to quark vs. anti-quark initial state PDFs. }
    \label{tab:F0}
}%
\end{table}
\begin{table}[]
{\scriptsize %
    \centering
    \begin{tabular}{|c|c|c|c|c|}
    \hline 
        Type  &  J & Q & s-channel process& coupling \\
 
        \hline 
        $V^L_0$ & 1  & +2/3 
        &  
        $e^+_R d_L \rightarrow  $
        \begin{tabular}{c} $l^+ d $ \\ $ \bar{\nu} _l u$\end{tabular}
        
        & \begin{tabular}{c} $\lambda _L$ \\ -$\lambda _L$\end{tabular} 
       \\
        \hline 
        
        $V^R_0$ & 1  & +2/3 
        &   $e^+_L d_R \rightarrow  \it{l}^+ d$
        &  $\lambda _R$   \\
         \hline 
        $\tilde{V}^R_0$ & 1  & +5/3 
        &   $e^+_L u_R \rightarrow \it{l}^+ u$
        &  $\lambda _R$  \\
        
       \hline 
        $V^L_1$ & 1  & \begin{tabular}{c} +2/3   \\ \\ +5/3\end{tabular} 
        &  
        \begin{tabular}{c} 
        $e^+_R d_L \rightarrow $ 
        \begin{tabular}{c} $l^+ d $ \\ $\bar{\nu} _l u$\end{tabular} \\
       $ e^+_R u_L \rightarrow \it{l}^+ u$
        \end{tabular} 
    
       & \begin{tabular}{c} -$\lambda _L$ \\ $\lambda$ \\ $\sqrt{2} \lambda_L$\end{tabular} \\


        \hline 
        \hline 
        $S^L_{1/2}$ & 0 & +5/3 
        & $e^+_R u_R \rightarrow  \it{l}^+ u$
        & $\lambda _L$ \\
        
        \hline 
         $S^R_{1/2}$ & 0 &  \begin{tabular}{c}+2/3  \\ +5/3 \end{tabular}
        & \begin{tabular}{c} $e^+_L d_L \rightarrow  \it{l}^+ d$ 
        \\  $e^+_L u_L \rightarrow  \it{l}^+ u$  \end{tabular}
        & \begin{tabular}{c} -$\lambda _R $ 
        \\   $\lambda _R$  \end{tabular}
     \\   
      
      \hline 
        $\tilde{S}^L_{1/2}$  & 0 & +2/3 
        &   $e^+_R d_R \rightarrow \it{l}^+ d$
        &  $\lambda _L$  \\
       \hline
        
    \end{tabular}
    }
    \caption{The F = 0 leptoquarks in the Buchm\"uller-R\"uckl-Wyler classification. For F=0 leptoquarks,  the s-channel process dominates with a positron beam due to quark vs. anti-quark initial state PDFs.}
    \label{tab:F2}

\end{table}

As shown schematically in Fig.~\ref{fig:LQtreelevel}, the LQs  mediate CLFV transitions at tree-level, allowing for larger cross sections compared to other scenarios in which CLFV processes are typically loop suppressed. 
For LQ masses $M_{LQ}\gg \sqrt{s}$, the tree-level processes in Fig.~\ref{fig:LQtreelevel} are described by contact interactions. In this approximation,  the cross-sections~\cite{Gonderinger_2010} for $e^-N \to \mu^-X$ via $F=0$ and $|F|=2$ LQs exhange take the form:

\begin{eqnarray}
\label{eq:elecF0LQcrosssection}
\sigma_{F=0}^{e^-p}&=& \sum_{\alpha,\beta} \frac{s}{32\pi} {\left[  \frac{\lambda_{1\alpha}\lambda_{2\beta}}{M_{LQ}^2}  \right]}^2 \int dx \int dy \\ 
&&\qquad \Big \{ x \bar{q}_\alpha(x,x s) f(y) + x q_\beta(x,-u) g(y)\>\> \Big \}, \nonumber \\ \nonumber
\end{eqnarray}

\begin{eqnarray}
\label{eq:elecF2LQcrosssection}
\sigma_{|F|=2}^{e^-p}&=& \sum_{\alpha,\beta} \frac{s}{32\pi} {\left[ \frac{\lambda_{1\alpha}\lambda_{2\beta}}{M_{LQ}^2} \right] }^2 \int dx \int dy \\
&&\qquad \Big \{ x q_\alpha(x,x s) f(y) + \bar{q}_\beta(x,-u) g(y) \>\>\Big \}. \nonumber 
\end{eqnarray}
Similarly, for $e^+N \to \mu^+X$, the $F=0$ and $|F|=2$ LQ exhange cross section takes the form:

\begin{eqnarray}
\label{eq:posF0LQcrosssection}
\sigma_{F=0}^{e^+p}&=& \sum_{\alpha,\beta} \frac{s}{32\pi} {\left[  \frac{\lambda_{1\alpha}\lambda_{2\beta}}{M_{LQ}^2}  \right]}^2 \int dx \int dy \\ 
&&\qquad \Big \{ x q_\alpha(x,x s) f(y) + x \bar{q}_\beta(x,-u) g(y)\>\> \Big \}, \nonumber \\ \nonumber
\end{eqnarray}
\begin{eqnarray}
\label{eq:posF2LQcrosssection}
\sigma_{|F|=2}^{e^+p}&=& \sum_{\alpha,\beta} \frac{s}{32\pi} {\left[ \frac{\lambda_{1\alpha}\lambda_{2\beta}}{M_{LQ}^2} \right] }^2 \int dx \int dy \\
&&\qquad \Big \{ x \bar{q}_\alpha(x,x s) f(y) + q_\beta(x,-u) g(y) \>\>\Big \}, \nonumber 
\end{eqnarray}
respectively. Here the kinematic variables $u=x(y-1)s$ and  $f(y)=1/2, g(y)=(1-y)^2/2$  for a scalar LQ and  $f(y)=2(1-y)^2, g(y)=2$ for a vector LQ.  The $\lambda_{ij}$ couplings are the lepton-quark-LQ couplings  where first and second indices denote the lepton and quark generations respectively, and can be related to the $h$ and $g$ couplings that appear at the Lagriangian level in Eqs.~(\ref{eq:F0Lag}) and (\ref{eq:F2Lag}), up to overall signs and factors of $\sqrt{2}$ which can be shown in the last columns of Tables~\ref{tab:F2} and \ref{tab:F0}, and the subscripts $L$ or $R$ denote left-handed or right-handed coupling of the LQ to lepton.
Note, that the first and second terms in the cross section formulae arise from an $s$-channel and $u$-channel LQ-exchange, respectively.

A global analysis using data obtained from the use of unpolarized and polarized electron and positron beams, as well as unpolarized and polarized nuclear targets, can allow for contraints on specific LQ states or combinations of states. Such an analysis can also be perfomed in the SMEFT framework~\cite{Boughezal:2021kla,Boughezal:2020uwq,Cirigliano:2021img}. In particular, the lepton beam polarization can be used to distinguish between contributions from left-handed and right-handed LQs. Comparing limits~\cite{Furletova:2018nci} obtained using a  positron beam with those obtained from an electron beam can also help untangle contributions from F=0 and $|$F$|$=2 LQs due to the different combinations of quark and anti-quark parton distribution functions (PDFs) that appear in the s- and u-channels, as seen in Eqs.(\ref{eq:elecF0LQcrosssection}-\ref{eq:posF2LQcrosssection}). Finally, the use of proton vs deutron nuclear targets can distangle contributions of the different electric charge states of the LQs corresponding to coupling to up or down type quarks.  Thus, the positron beam studies can be complementary to CLFV studies planned with an electron beam at the SOLID~\cite{chen2014white} experiment at JLAB and at the proposed Electron-Ion collider (EIC)~\cite{Accardi:2012qut,Gonderinger_2010}.

\section{CLFV Limits}
The HERA~\cite{H1:limit,HERA-ZEUS} collaborations quantified the results of the CLFV searches by setting limits on the coupling to mass ratios \\

\begin{eqnarray}
   \chi_{\alpha \beta} \equiv \frac{\lambda_{1\alpha}\lambda_{2\beta}}{M_{LQ}^2}, \\ \nonumber
\end{eqnarray}
that appear in the cross sections in Eqs.~(\ref{eq:elecF0LQcrosssection}-\ref{eq:posF2LQcrosssection}). For example, for the F=0 LQ state
$S_{1/2}^L$, limits of $\chi_{1 1} < 0.6$ TeV$^{-2}$ and $\chi_{1 2} < 0.7$ TeV$^{-2}$ were found~\cite{H1:limit}. A complete listing of HERA limits on various LQ states can be found in Refs.~\cite{H1:limit,HERA-ZEUS}. For the purposes of comparing the reach at CEBAF to HERA limits, it becomes useful to define the quantity~\cite{Gonderinger_2010}\\

\begin{eqnarray}
\label{eq:zHERA}
   z\equiv \frac{\chi_{\alpha \beta}}{\chi_{\alpha \beta}^{\>\>\rm HERA}}, \\ \nonumber
\end{eqnarray}
\\
which gives the ratio of $\chi_{\alpha \beta}$ to its upper limit,  $\chi_{\alpha \beta}^{\rm HERA}$, as set by HERA~\cite{H1:limit,HERA-ZEUS}. Thus, the cross sections in Eqs.~(\ref{eq:elecF0LQcrosssection}-\ref{eq:posF2LQcrosssection}) can be written as a function of the variable $z$. The cross section at $z=1$ corresponds to using evaluating it at the HERA limit $\chi_{\alpha \beta}=\chi_{\alpha \beta}^{\rm HERA}$. Similarly, $z<1$ corresponds to evaluating the cross section below the HERA limit $\chi_{\alpha \beta}<\chi_{\alpha \beta}^{\rm HERA}$.

A positron beam at CEBAF can improve on the HERA limits. The HERA collider operated with a center of mass energy $\sqrt{s}=300$ GeV, much bigger than $\sqrt{s}\sim 4.5$ GeV for the CEBAF facility. Thus, for a fixed value of $\chi_{\alpha \beta}$, the LQ cross sections in Eqs.~(\ref{eq:elecF0LQcrosssection}-\ref{eq:posF2LQcrosssection}) at CEBAF are expected to be smaller by a factor of $\sim (4.5/300)^2 =2.25\times 10^{-4}$ compared to HERA. However, compared to HERA, the CEBAF facility will have an instantaneous luminosity that will be larger by a factor of $\sim 10^{6}$ or $10^7$.  Running the CEBAF experiment with instantaneous luminosity ${\cal L}\sim 10^{38}$ cm$^{-2}$ s$^{-1}$ for five years will yield the integrated luminosity ${\cal L}_{\rm int.}\sim 5\times 10^6$ fb$^{-1}$.  Without taking efficiencies into account, this will allow for sensitivity to cross sections as small as $\sigma \sim 0.2 \times 10^{-6}$ fb which will yield a number of events of order one. 

\begin{figure}
\includegraphics[width=1.0\linewidth]{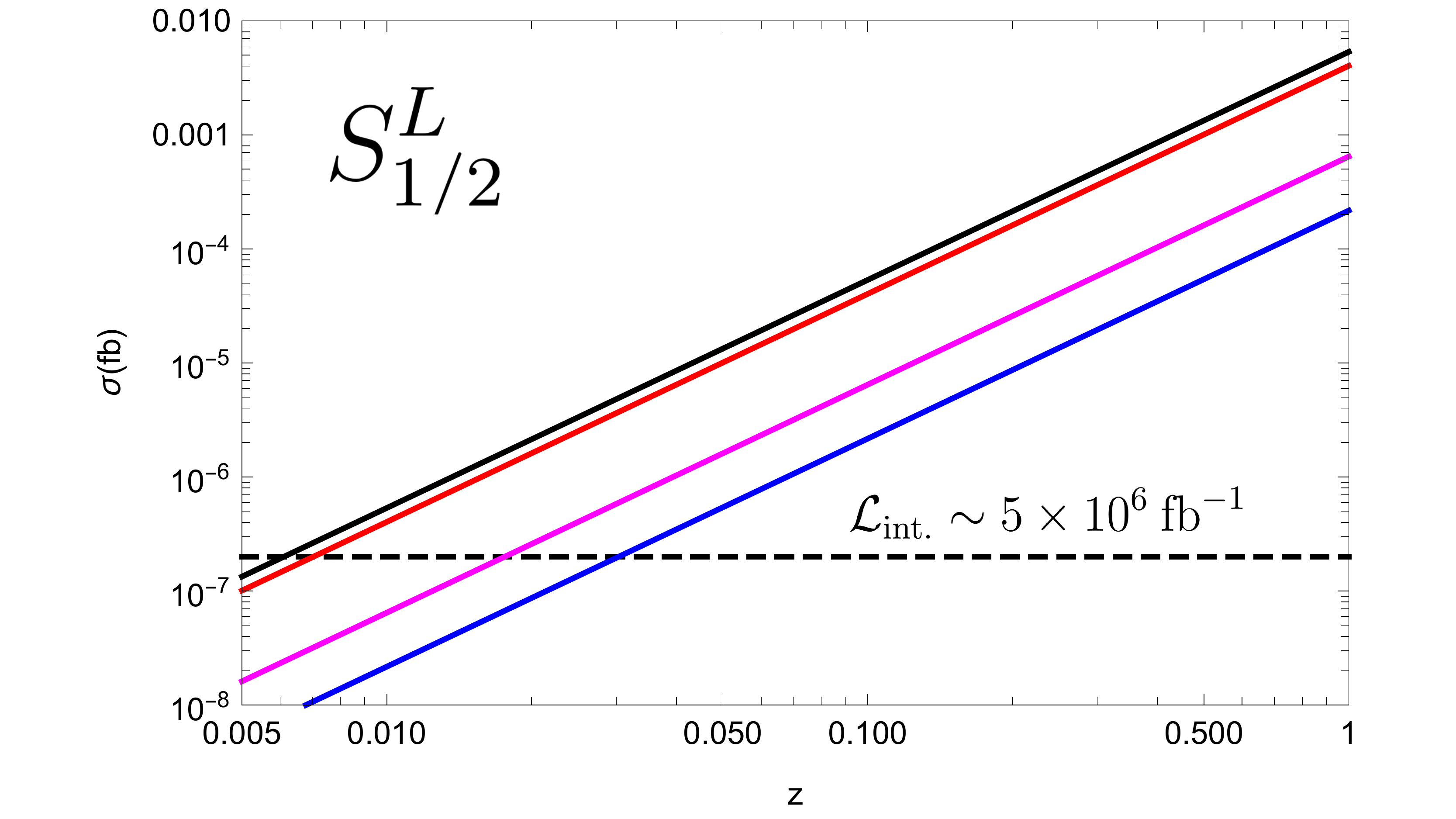}
\caption{The cross section for $e^+N\to \mu^+ X$ with center of mass energy $\sqrt{s}=4.5$ GeV, via exchange of the F=0 scalar LQ, $S_{1/2}^L$,  as a function of the ratio $z$ defined in Eq.~(\ref{eq:zHERA}). The red, black, magenta, and blue solid lines correspond to the choices $(\alpha,\beta)=\{11,12,21,22\}$ in Eq.~(\ref{eq:posF0LQcrosssection}) with all other terms set to zero. An integrated luminosity of ${\cal L}\sim 5\times 10^6$fb$^{-1}$ will allow sensitivity to cross sections as small as $\sigma \sim 0.2 \times 10^{-6}$ fb (horizontal dashed line).}
\label{fig:LQzcrosssection}
\end{figure}

In Fig.~\ref{fig:LQzcrosssection}, we show the cross section at CEBAF for $e^+N\to \mu^+ X$, via the exchange of the F=0 left-handed scalar LQ, $S_{1/2}^L$, as a function of $z$. The various lines correspond to the cross section arising for a specific choice of $(\alpha,\beta)$ in Eq.~(\ref{eq:posF0LQcrosssection}), with all other terms set to zero. The set of four choices  $(\alpha,\beta)=\{11,12,21,22\}$ correspond to the red, black, magenta, and blue colors, respectively. We see that sensitivity to a cross section $\sigma \sim 0.2 \times 10^{-6}$ fb, will translate into a limit in the range $z\sim [0.005-0.05]$, depending on the specific choice of $(\alpha, \beta)$ corresponding to an improvement by two or orders of magnitude over the HERA limits, corresponding to $z=1$. 

The expected improvement on the HERA limits can also be complementary to the more stringent limits coming from other low energy experiments. For example, searches~\cite{SINDRUM} of $\mu-e$ conversion on gold nuclei yield the constraint, $CR(\mu-e, Au) = \frac{\Gamma(\mu^- Au \to e^- Au)}{\Gamma_{\rm capture}}  
 < 7.0\times 10^{-13}$.
Since this $\mu-e$ conversion involves the $Au$ nucleus in the initial and final state, it only constrains the product of couplings $\lambda_{1\alpha}\lambda_{2\beta}$ that both involve only same quark generation ($\alpha=\beta$). This yields constraints on $\chi_{11}$ and $\chi_{22}$ that are much more stringent than the HERA limits. For example,  the corresponding limits from $\mu-e$ conversion are $\chi_{11}^{\mu-e} \sim 5.2 \times 10^{-5}$ TeV$^{-2}$ and $\chi_{22}^{\mu-e} \sim 9.4 \times 10^{-4}$ TeV$^{-2}$.  This can be contrasted with the HERA limits for the $S_{1/2}^L$ LQ which are $\chi_{11}^{\rm HERA} \sim 0.6$ TeV$^{-2}$ and $\chi_{22}^{\rm HERA} \sim 2.4$ TeV$^{-2}$. Thus, the expected  improvement at CEBAF over the HERA limits is still not enough to compete with the constraints from $\mu-e$ conversion. However, $\mu-e$ conversion does not constrain $\chi_{12}$ which involves quarks from both the first and second generations and HERA in fact gives the best limit for $S_{1/2}^L$. Thus, CEBAF can yield significant improvement in the region of the theory that might not be accessible to other low energy experiments. Similarly, for some other LQs, such as $\tilde{S}_{1/2}^L$, which differs from  $S_{1/2}^L$ in hypercharge, more stringent limits of $\chi_{12}\sim 2\times 10^{-5}$ TeV$^{-2}$, come from searches of the CLFV kaon decays $K\to \mu^- e^+$~\cite{H1:limit}. However, once again, while CLFV kaon decays constrain the $\tilde{S}_{1/2}^L$ which couples to anti-leptons and down-type quarks, it does not constrain $S_{1/2}^L$ which couples to anti-leptons and up-type quarks. 

Similarly, much stronger constraints are expected from CLFV searches at the Large Hadron Collider (LHC) ~\cite{PDG:2020}. However, compared to the LHC evnironment, a polarized lepton beam in the initial state allows better control in isolating effects from different types of LQs.

A lepton beam polarization can allow one to distinguish between left-handed and right-handed LQ effects. The lepton beam polarization is defined as: 

\begin{eqnarray}
P_e = \frac{N_R-N_L}{N_R+N_L}
\end{eqnarray}
\\
where $N_R$ and $N_L$ denote denote the number of right-handed and left-handed leptons (electrons or positrons). Correspondingly, the chiral coupling of the LQ states to the lepton beam leads to cross section having a linear dependence on the beam polarization:

\begin{eqnarray}
\label{eq:LinearPol}
\sigma(P_e) = (1 \pm P_e) \sigma(P_e=0). \\ \nonumber
\end{eqnarray}

\begin{figure}
\includegraphics[width=1.0\linewidth]{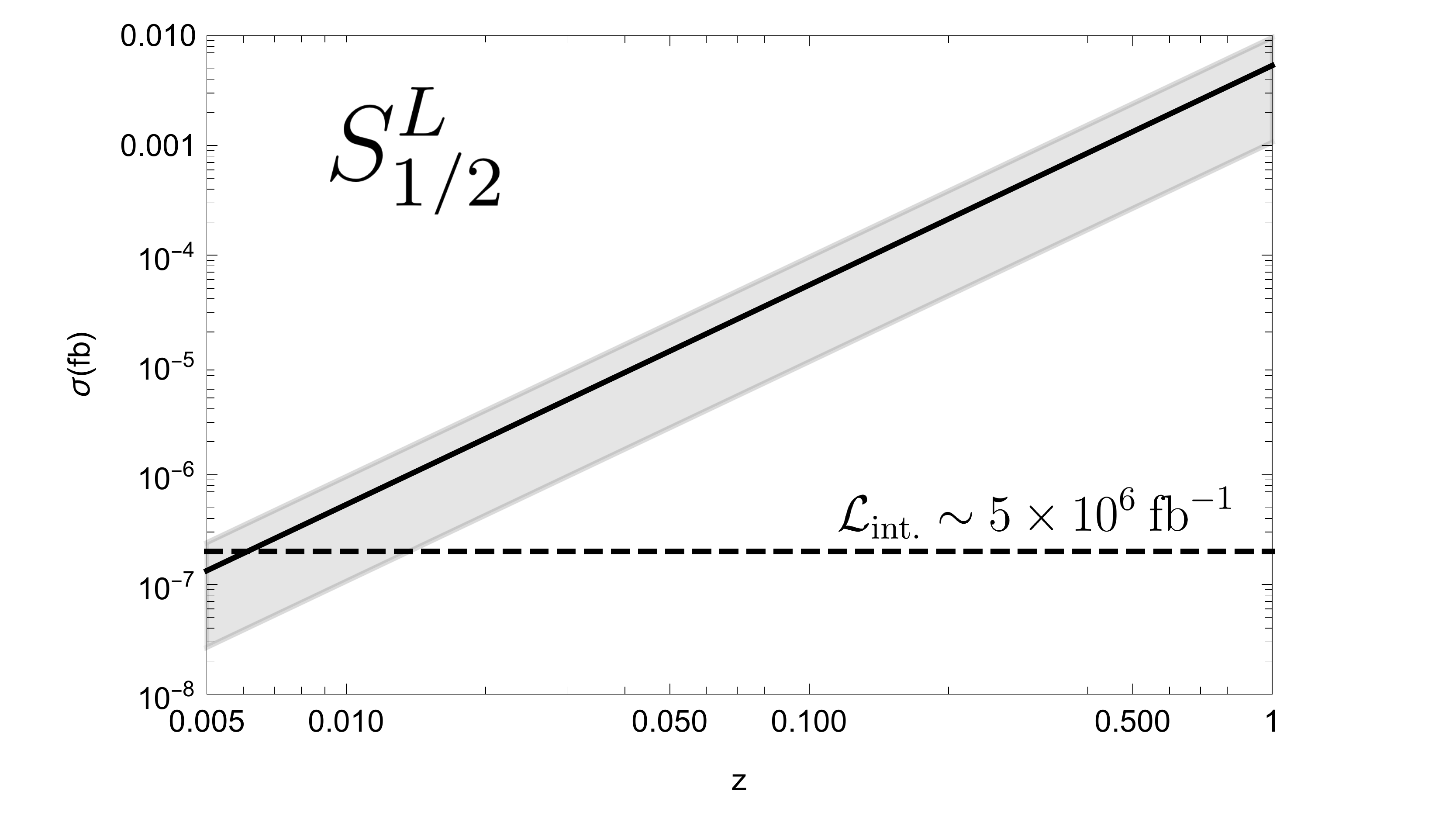}
\caption{The positron beam polarization dependence of cross section for $e^+N\to \mu^+ X$ with center of mass energy $\sqrt{s}=4.5$ GeV, via exchange of the F=0 scalar LQ, $S_{1/2}^L$,  as a function of the ratio $z$ defined in Eq.~(\ref{eq:zHERA}). The solid black line corresponds to the cross section for an unpolarized positron beam ($P_e=0$). The gray band corresponds to the linear variation of the cross section with beam polarization, as shown in Eq.~(\ref{eq:LinearPol}). The size of the band corresponds to a variation of the beam polarization between [-80\%,80\%].}
\label{fig:LQzPolcrosssection}
\end{figure}

Thus, by varying the degree of lepton beam polarization, one can better constrain left-handed and right-handed LQ states. In Fig.~\ref{fig:LQzPolcrosssection}, we show the effect beam polarization when it is varied over the range $P_e=[-80\%$, $80\%]$, according to Eq.~(\ref{eq:LinearPol}). The solid black line denotes the unpolorized cross section $\sigma(P_e=0)$ for the $S^L_{1/2}$ LQ state with $\lambda_{11}\lambda_{22}$ non-zero and all other LQ couplings set to zero. In terms of $\chi_{12}$, the HERA limit is $\chi_{12}^{\rm HERA}\sim 0.7$ TeV$^{-2}$. The gray band around the solid black line corresponds to the variation of the cross section with polarization.

The CLFV studies at CEBAF will also complement future studies at the Electron-Ion Collider (EIC) which will also search for $e\to \tau$ CLFV transitions~\cite{Accardi:2011mz,Gonderinger_2010,boer2011gluons}. In fact, due to its much larger luminosity, the CEBAF bounds on CLFV transitions between the first two lepton generations are still expected to be stronger than at the EIC.  CLFV at CEBAF will also complement planned studies~\cite{Gninenko:2018num} using electron and muon beams in the NA64 experiment. Thus, in general, the CEBAF positron program to explore CLFV processes can provide new insights and be complementary to other searches across a wide variety of experiments.


\section*{Conclusions}
A polarized positron beam at CEBAF can play an important role in the search for charged lepton flavor violation, through a search for the process $e^+N \rightarrow \mu^+X$,  at the intensity frontier.  The polarization of the positron beam can distinguish between different CLFV mechanisms, such as left-handed vs. right-handed Leptoquarks.  It's large luminosity allows for improving on HERA limits by two or three orders of magnitude and complementing CLFV searches in other experiments, including proposed CLFV studies at the Electron-Ion Collider (EIC) via searches for  $eN\rightarrow \tau X $~\cite{Accardi:2011mz,Gonderinger_2010,boer2011gluons} and the NA64 experiment~\cite{Gninenko:2018num}.

\section*{Acknowledgment}
This material is based upon work supported by the U.S. Department of Energy, Office of Science, Office of Nuclear Physics under contract DE-AC05-06OR23177.

\bibliographystyle{h-physrev3.bst}
\bibliography{CLFV}


%
%
%
%

\end{document}